\documentclass[pra, onecolumn]{revtex4}
\usepackage{graphicx}

\usepackage{ulem}
\usepackage[section]{placeins}
\usepackage{amsmath}
\usepackage{amssymb}
\usepackage{dsfont}
\usepackage{natbib}

\usepackage{subfiles}
\usepackage{graphicx}
\usepackage{array}
\usepackage{calrsfs}
\usepackage{hhline}
\usepackage{relsize}

\begin{document}

\title{Hong-Ou-Mandel interference with a diode-pumped 1-GHz Ti:sapphire laser}

\author{Imogen Morland, Hanna Ostapenko, Feng Zhu, Derryck T. Reid and Jonathan~Leach$^{*}$}
\address{Institute of Photonics and Quantum Sciences, Heriot-Watt University, David Brewster Building, Edinburgh EH14 4AS, UK}
\address{$^*$j.leach@hw.ac.uk}


\begin{abstract}

Correlated photon pairs generated through spontaneous parametric down-conversion (SPDC) are a key resource in quantum optics. In many quantum optics applications, such as satellite quantum key distribution (QKD), a compact, high repetition rate pump laser is required. Here we demonstrate the use of a compact, GHz-rate diode-pumped three-element Kerr-lens-modelocked Ti:sapphire laser for the generation of correlated photon pairs at 790 nm. We verify the presence of indistinguishable photons produced via SPDC using Hong-Ou-Mandel (HOM) interferometry and observe a dip in coincidence counts with a visibility of 81.8\%. 
   
\end{abstract}
\maketitle

\section{\label{sec:Intro}Introduction}

Hong-Ou-Mandel (HOM) interference \cite{hong1987measurement} is a two-photon effect that demonstrates the quantum nature of indistinguishable single photons. It has applications in quantum optics \cite{shih1988new,rarity1990two, santori2002indistinguishable}, quantum communication \cite{ekert1991quantum,yin2017satellite} and quantum computing \cite{kok2007linear}. HOM interference also plays an important role in quantum metrology, where quantum phenomena such as entanglement are used \cite{boto2000quantum,giovannetti2011advances}. For example, the generation of photon pairs via spontaneous parametric down-conversion (SPDC) can be used as a source for N00N states \cite{dowling2008quantum}. These states are necessary to achieve the fundamental quantum limit for phase sensitivity known as the Heisenberg limit \cite{bouchard2020two}. HOM interference also plays a key role in Bell-state measurements \cite{weinfurter1994experimental}, which are used in entanglement swapping \cite{zhang2016engineering,zhang2017simultaneous} and quantum teleportation \cite{bouwmeester1997experimental}. Furthermore,  quantum optical coherence tomography (QOCT) \cite{nasr2003demonstration} builds on optical coherence tomography (OCT) \cite{huang1991optical} by using HOM interferometry to measure the depth profile of biological samples to micrometre resolution \cite{abouraddy2002quantum, nasr2004dispersion, lopez2012quantum}. HOM interferometry has been used in the high precision measurement of time delays \cite{dauler1999tests,branning2000simultaneous} and polarization \cite{harnchaiwat2020tracking} and when combined with statistical estimation theory, nanometer precision has been achieved \cite{lyons2018attosecond}. More recently, HOM interferometry was used to image micrometre-scale depth features \cite{ndagano2022quantum}.

A photon pair source with high brightness, visibility and collection efficiency combined with a high generation rate is required in many quantum optics applications \cite{bouwmeester1997experimental, liao2017satellite}. High brightness integrated sources have been demonstrated  \cite{vergyris2017fully, fan2007bright, sansoni2017two} and there are many examples where the brightness, visibility and collection efficiency are close to optimal \cite{zhong201812, meyer2018high}. A domain-engineering technique for tailoring the crystal nonlinearity to generate indistinguishable and spectrally pure photons without filtering has also been demonstrated \cite{graffitti2018independent}.

The development of GHz ultrafast lasers has opened the door for high generation rate photon pair sources, and such GHz lasers have been used to generate SPDC at telecommunication wavelengths \cite{zhang2008generation, jin2014efficient, ngah2015ultra}. Recently, the HOM interference between two photons from two independent 10 GHz sources, spaced by distances of up to 100 km has been demonstrated with visibilities in excess of 90\% \cite{d2020universal}. Further work has been done to improve the indistinguishability of photon pairs produced via SPDC with a GHz laser  using temporal filtering techniques \cite{miyanishi2020robust}. The effect of GHz repetition rate lasers on the spectral purity has been modelled and confirmed experimentally by demonstrating high-visibility HOM interference between two independent heralded single photons (HSPs) generated by SPDC with $3.2$ GHz pump pulses \cite{tsujimoto2021ultra}.

One of the issues with the generation of quantum states of light is that typical ultrafast lasers are bulky and expensive. Here we demonstrate the use of a compact, low cost GHz laser \cite{ostapenko2022three, ostapenko2023design}, with characteristics compatible for applications where size, weight and power are important, such as satellite quantum key distribution (QKD) \cite{liao2017satellite}. We show that this compact GHz laser can be used to produce indistinguishable photon pairs for quantum optics as demonstrated by the observation of a high visibility HOM dip.

\section{\label{sec:Set up}Experimental Set-up}

\begin{figure*}[ht]
	\centering
	\includegraphics[width=\textwidth]{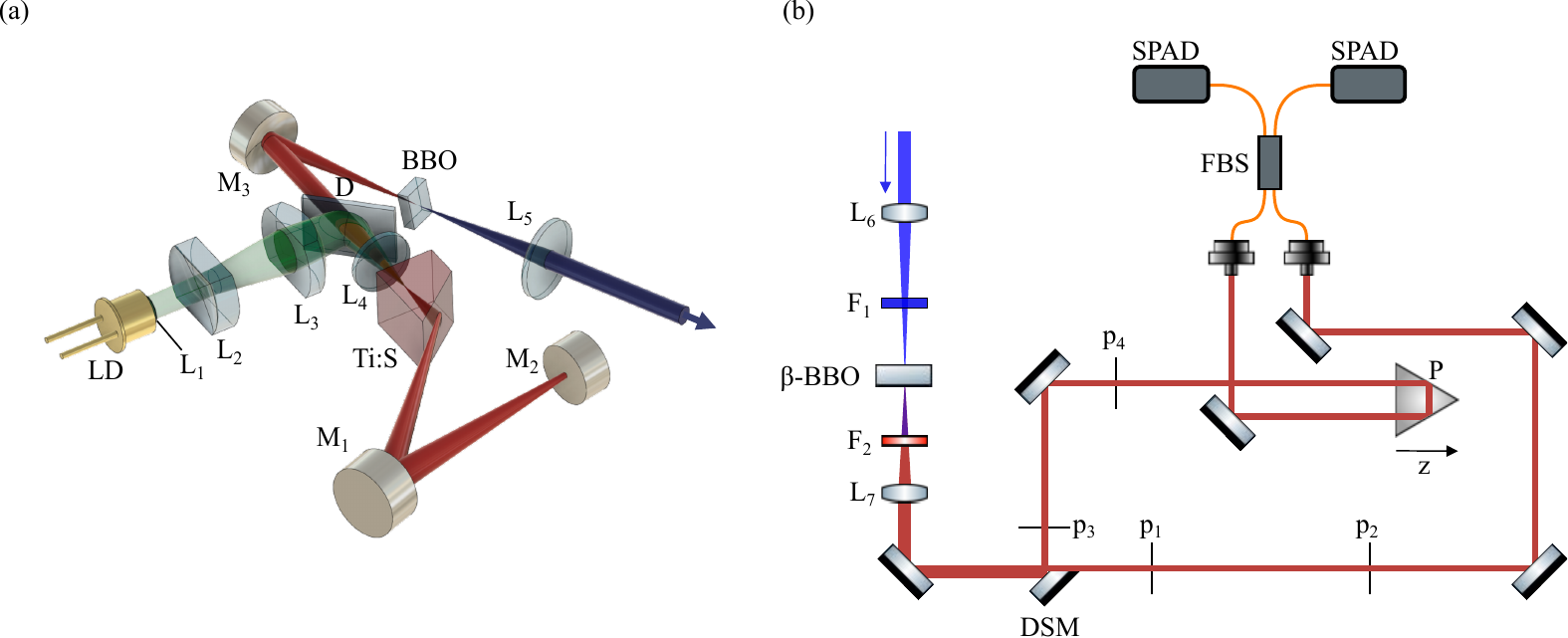}
	\caption{The experimental setup has two parts: the laser cavity and the Hong–Ou–Mandel (HOM) experiment. The output beam in (a) is the input beam in (b). (a) The laser cavity was pumped by a 1.1 W laser diode (Nichia NDG7D75), which is collimated via three lenses ($L_1$, $L_2$ and $L_3$), and focused onto the Ti:sapphire gain crystal using a plano-convex achromatic doublet lens with a focal length of 45 mm ($L_4$). The oscillator includes a concave mirror ($M_1$) with R = -50 mm and a plane end mirror ($M_2$). The beams are separated using a dichroic mirror (D) and focused into a $\beta$-BBO with a concave mirror $M_3$ with R = -75 mm and 395 nm pulses are produced. A series of lenses and cylindrical lenses were used to collimate the beam and reduce the beam radius. (b) The up-converted light was filtered using a bandpass filter, $F_1$ (Thorlabs FBH405-10) and then focused into a second Type I $\beta$-BBO and the signal and idler photons were imaged via lenses ($L_6$ and $L_7$) with focal lengths $f_1~ = 150$ mm and $f_2~ = 60$ mm. The signal and idler photons were filtered using a tilted 790 nm bandpass filter $F_2$ (Thorlabs FBH790-10) and directed down different paths using a D shaped mirror (DSM) and two pinholes were placed in each of the paths. The down-converted photons are coupled into a SM FBS, each with lenses of focal length 7.5 mm, via mirrors and a prism (P). Single photon counts were recorded by each SPAD (Excelitas SPCM-NIR), and from this coincidences counts are calculated as a function of the translation stage position, z.}
	\label{Fig: Exp}
\end{figure*}
Our generation and measurement system comprises three stages. In the first stage, a laser diode was used to pump the three-element Kerr-lens-modelocked Ti:sapphire laser and 790 nm, 105 fs pulses were produced at 1 GHz. Next, these laser pulses were frequency doubled using a type I $\beta$-BBO crystal to produce 395 nm light. Then, a second $\beta$-BBO crystal was used to produce photon pairs via SPDC, which traveled through each arm of the HOM interferometer and were coupled into a single-mode fiber beam splitter (SM FBS). Coincidence counts were then recorded by single-photon avalanche diodes (SPADs) as a function of the relative optical delay between the photon pairs. The visibility of the dip in coincidences confirms the generation of indistinguishable single photons, when it exceeds 50\%.

The laser platform used for the experiment was a diode pumped three-element Kerr-lens-modelocked Ti:sapphire cavity \cite{ostapenko2022three, ostapenko2023design} as shown in Fig. \ref{Fig: Exp} (a). A single laser diode (Nichia NDG7D75) operated at a maximum pump power of 1.1 W was used to pump the cavity. The fast axis of the diode was collimated using $L_1$ and the slow axis was expanded using two cylindrical lenses ($L_2$ and $L_3$), which act as a 1:6 telescope. 

The Ti:sapphire crystal has an absorption of 4.1 cm$^{-1}$ at 532 nm, a figure of merit >200, and a plane-Brewster geometry where the plane side of the mirror is coated to be transmitting at 450-530 nm and 99\% reflective at 770-830 nm. The coating specifications allow the plane side of the crystal to act as both the pump in-coupling mirror for the cavity and the output coupler of 1\%. A focusing achromatic doublet lens, $L_4$, with 45 mm focal length was used to both focus the pump beam into the crystal and collimate the laser output, and the two beams were separated by a dichroic mirror (D) outside the cavity. 

The oscillator itself also includes a concave mirror with R = -50 mm and a plane end mirror with a Gires-Tournois interferometer (GTI) coating of -550 fs$^2$ used for dispersion optimisation of the cavity. The 790 nm output beam was then separated from the pump using dichroic mirror D and focused into a 2 mm thick $\beta$-BBO crystal with a focusing mirror $M_3$ with R = -75 mm and second-harmonic generation (SHG) pulses were produced. A series of lenses and cylindrical lenses collimated the beam and reduced it's radius. 

The HOM experiment is shown in Fig \ref{Fig: Exp} (b). The up-converted pulses were filtered using a bandpass filter angle-tuned to 395 nm with a full width at half maximum (FWHM) of 10 nm (Thorlabs FBH405-10), $F_1$, and the beam was focused into another Type I $\beta$-BBO via lens $L_6$ with focal length $f_1~ = 150$ mm. In the alignment stage, the photon pairs generated via SPDC were relayed using an electron multiplying charge-coupled device (EMCCD) and the angle of the $\beta$-BBO crystal was adjusted to give the desired phase matching condition. Two pinholes were then positioned in each arm of the HOM interferometer such that the signal and idler photons were selected and an infrared (IR) laser diode was used for back projection. This was done using two alignment mirrors in each arm such that the light from the IR laser diode propagated through the pinholes. The coincidence counts were coupled into single-mode (SM) fibers using 7.5 mm focal length lenses and were optimized with alignment mirrors and the three axis fiber alignment stages. A 650 nm long pass filter was used to maximise the signal and achieved a quantum contrast \cite{zhu2021high} of 1800 and collection efficiencies of 10.9\% and 9.6\% for each SPAD. 

Once aligned, the long pass filter was replaced with a 790 nm bandpass filter, $F_2$ (Thorlabs FBH790-10) with a full width at half maximum (FWHM) of 10 nm, narrowing the spectrum in the frequency domain in order to broaden the width in the time domain. The photon pairs were imaged using another lens ($L_7$) with $f_2~ = 60$ mm. The SM fibers were connected to a SM FBS which was connected to two SPADs (Excelitas SPCM-NIR), each recording singles counts. The SPADs are enhanced for detecting in the near infrared (NIR) region, with a peak photon detection efficiency (PDE) of 70\% for 780 nm wavelengths. The time tags of the photon arrival time were recorded using a PicoQuant TimeHarp 200 allowing coincidences to be established between the two detectors within a ~1 ns window. 

The second arm contains a retroreflector prism (P) mounted on a motorised translation stage (CONEX-MFACC), which was scanned to find the dip in coincidences counts. When the location of the dip was found, fiber polarization controllers were rotated to match the polarization of the two photons by minimising the dip in coincidence counts at the dip location. Once the coincidences counts have been maximised outside the dip and minimised inside the dip, the coincidence data as a function of stage position can be recorded.

\begin{figure}[ht]
\centering
\centerline{\includegraphics[]{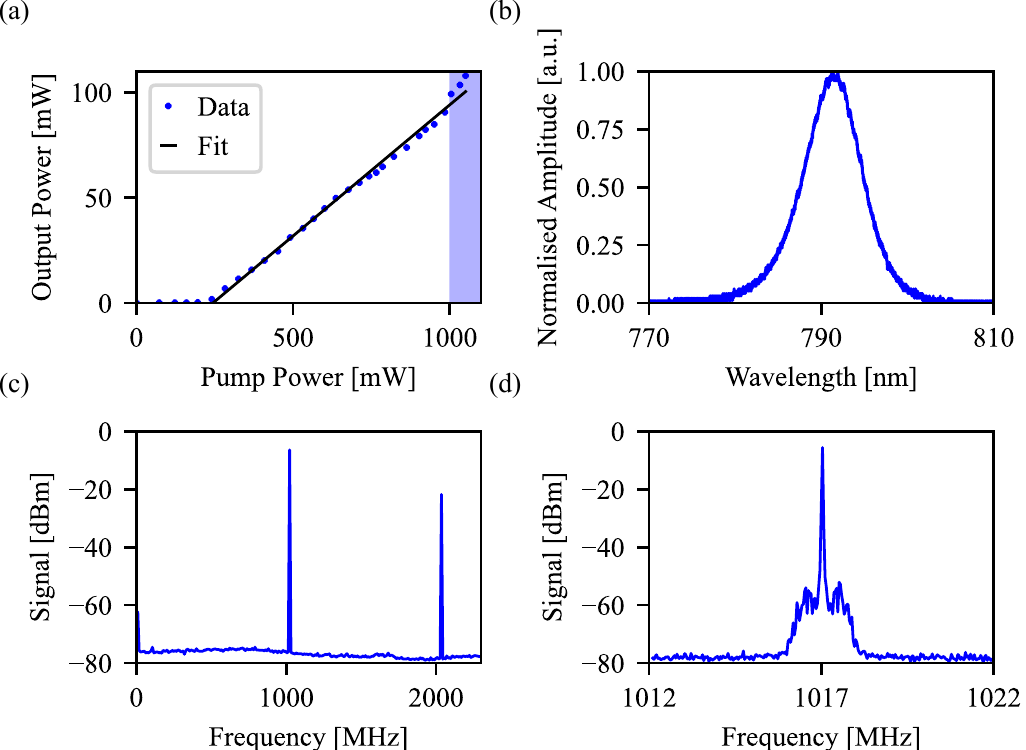}}
\caption{ (a) The output power as a function of pump diode power, with a linear fit showing a slope
efficiency of 12.4\%. Self-starting Kerr-lens-modelocking occurs for pump powers greater than 1005 mW as indicated by the blue shaded region. (b) Modelocked spectra recorded for pump powers of 1100 mW with a central wavelength of 791 nm and FWHM of 8.1 nm. (c) Radio-frequency spectrum recorded at the instrument-limited resolution bandwidth of 15 kHz. The repetition rate is centered at 1.02 GHz. (d)Detail of the fundamental peak of the radio-frequency spectrum shown in (c).
\label{Fig: Laser Results}}
\end{figure}

The characteristics of the diode pumped three-element Kerr-lens-modelocked Ti:sapphire cavity are shown in Fig.~\ref{Fig: Laser Results}. The output power as a function of the pump power is shown in Fig.~\ref{Fig: Laser Results} (a) and a linear fit implied the slope efficiency of the laser was 12.4\%. Self-starting Kerr-lens-modelocking occurs for pump powers greater than 1005 mW, as indicated by the shaded region. The normalised  modelocked spectrum is shown in Fig. \ref{Fig: Laser Results} (b) and has a central wavelength of 791 nm and a FWHM of 8.1 nm. Finally, the radio-frequency spectrum was recorded with an instrument-limited resolution bandwidth of 15 kHz and is shown in Fig. \ref{Fig: Laser Results} (c). A zoomed in view of the radio-frequency spectrum is shown in Fig. \ref{Fig: Laser Results} (d) and the repetition rate of the modelocked laser is found to be 1.02 GHz.

\section{\label{sec:Result}Results}

\begin{figure}[ht]
\centering
\centerline{\includegraphics[]{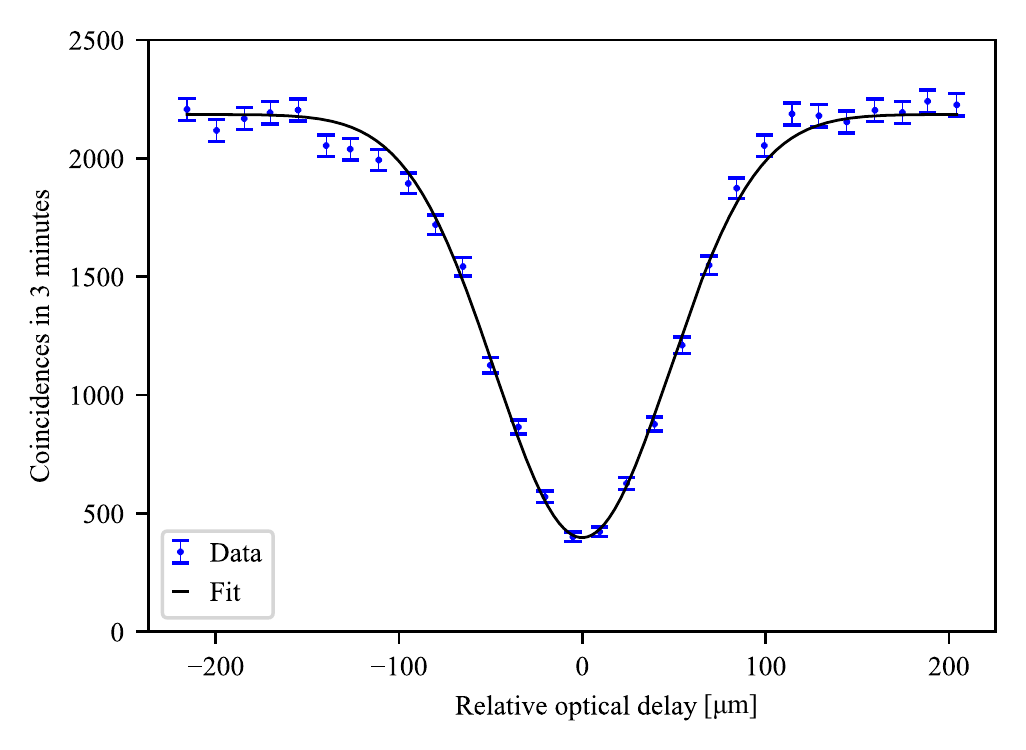}}
\caption{ HOM interference between signal and idler photons. A HOM dip is observed with a visibility of 81.8\% and a FWHM of $110~\mu m$. The error bars are calculated using Poisson statistics. The solid line is a Gaussian fit.
\label{Fig: Results}}
\end{figure}

Using this custom built Ti:sapphire modelocked laser (see \cite{ostapenko2022three, ostapenko2023design} for more information), we were able to observe a HOM dip with a visibility of 81.8\%, as shown in Fig. \ref{Fig: Results}. Coincidences were recorded for three minutes for every measurement and the FWHM of the dip was $110~\mu $m. In order to maximise the visibility of the HOM dip, the photons must be indistinguishable when interfering at the FBS. That is to say, that if the two photons have the same polarization and spectral properties and arrive at the FBS at the same time, the visibility will approach 100\%. To further improve the observed visibility, we would require more signal power to see changes to the coincidences inside the dip in real time.

\section{\label{sec:Conclusion}Conclusion}

In this work, we generate single photon pairs via SPDC using a compact, diode-pumped three-element Kerr-lens-modelocked Ti:sapphire cavity with a 1-GHz repetition rate  \cite{ostapenko2022three, ostapenko2023design}. We verify the presence of indistinguishable photon pairs using HOM interferometry by recording coincidence counts with SPADs and observe a dip in coincidences with a visibility of 81.8\%. These results show that we were able to create a quantum source of indistinguishable photon pairs using a simple and compact GHz-rate laser, opening new opportunities for GHz rate SPDC generation and detection in systems with low size, weight and power. 

We could achieve GHz-rate SPDC generation by removing the upconversion process described above, down-converting the 790 nm beam and detecting single photon pairs using superconducting nanowire single-photon detectors (SNSPDs). These detectors have a higher quantum efficiency and detection rate compared to SPADs, with quantum efficiencies of 98\% reported 
for detection at 1550 nm wavelengths \cite{reddy2020superconducting} compared to a 70\% quantum efficiency at 780 nm wavelengths for the SPADs used in this experiment. The elimination of the up-conversion process would reduce the loss in laser power and increase the probability of producing a photon pair in the SPDC process. 

In conclusion, we have generated indistinguishable photon pairs using a GHz rate diode-pumped laser, which is compact and low cost in comparison to typical ultrafast lasers used in SPDC to date. These characteristics make such a laser source a prime candidate in applications such as satellite QKD, where the size, weight and power of the laser are critical. 

\section*{Funding}
Science and Technology Facilities Council (ST/S505407/1, ST/V00040/1); Engineering and Physical Sciences Research
Council (EP/S001638/1, EP/T00097X/1, EP/P005446/1).

\section*{Disclosures}

The authors declare no conflicts of interest.

\end{document}